\documentclass{article}
\usepackage{spconf}
\usepackage{amsmath}
\usepackage[dvipdfmx]{graphicx}
\usepackage{amssymb}
\usepackage{multirow}
\usepackage{threeparttable}
\usepackage{cite}
\usepackage{booktabs}
\usepackage{makecell}
\usepackage{diagbox}

\def\Vec#1{\textit{\boldmath $#1$}}

\def\Smallsum{\mathop{\textstyle \sum}}

\title{Metric learning with background noise class\\ for few-shot detection of rare sound events}

\makeatletter
\def\@name{ \emph{Kazuki Shimada, Yuichiro Koyama, Akira Inoue}}
\makeatother

\address{Sony Corporation, Tokyo, Japan}
\begin{document}
\ninept

\maketitle

\begin{abstract}

Few-shot learning systems for sound event recognition have gained interests
 since they require only a few examples to adapt to new target classes without fine-tuning.
However,
 such systems have only been applied to chunks of sounds for classification or verification.
In this paper,
 we aim to achieve few-shot detection of rare sound events,
 from query sequence that contain not only the target events
 but also the other events and background noise.
Therefore,
 it is required to prevent false positive reactions to both the other events and background noise.
We propose metric learning with background noise class for the few-shot detection.
The contribution is to present the explicit inclusion of background noise as an independent class,
 a suitable loss function that emphasizes this additional class,
 and a corresponding sampling strategy that assists training.
It provides a feature space where the event classes and the background noise class are sufficiently separated.
Evaluations on few-shot detection tasks, using DCASE 2017 task2 and ESC-50,
 show that our proposed method outperforms metric learning without considering the background noise class.
The few-shot detection performance is also comparable to that of the DCASE 2017 task2 baseline system,
 which requires huge amount of annotated audio data.

\end{abstract}


\section{INTRODUCTION}
\label{sec:intro}

Detection of rare sound events
 has been regarded as an import role in many applications,
 such as monitoring~\cite{rouas2006audio}, audiovisual search~\cite{lew2006content,liu2016event},
 and sound command recognition~\cite{alvarez2019end}.
It is a common challenge to train a system to be robust,
 especially in the real-world scenarios where the target events only happen occasionally.
Recent competitions such as DCASE Challenge show
 great progress in detection of rare sound events
 using deep learning techniques~\cite{mesaros2017dcase,cakir2017convolutional,lim2017rare,kao2018r,phan2018weighted,arora2019deep,jati2019hierarchy}.
These studies are based on supervised learning,
 which requires huge amount of annotated audio data for training.
On the other hand, in these days, few-shot learning gains a lot of interests,
 which enables a system to adapt to the target classes with only a few examples~\cite{koch2015siamese,vinyals2016matching,snell2017prototypical}.
It can potentially reduce annotation costs of the training data.
In addition,
 it can also realize an event detection system that can be configured by giving a few samples of a target sound,
 such as an entrance bell or kitchen timer, by the users.

A few-shot learning systems can quickly adapt to unseen classes with only a few examples~\cite{koch2015siamese,vinyals2016matching,snell2017prototypical}.
There are two types of tasks; few-shot classification and few-shot verification.
Few-shot classification is given $k$ examples of new classes not seen during training,
 then classifies a query into one of these new classes.
For this task, deep metric learning is commonly used,
 which aims to learn representations that retain the class neighborhood structure
 so that similarity can be measured as a distance in the learned feature space~\cite{wang2014learning,hoffer2015deep,koch2015siamese,vinyals2016matching,snell2017prototypical}.
Besides these methods, which are proposed for computer vision,
 several metric learning methods have been proposed for tasks in acoustic signal processing.
Chou~{\it et al.} have tackled few-shot audio clip-level event classification~\cite{piczak2015esc}
 and introduce an attentional similarity module
 especially to classify related short sound events~\cite{chou2019learning}.
On the other hand,
 few-shot verification is given $k$ examples of a target class,
 then verifies whether a query sample is a target class or not.
Wang~{\it et al.} have worked on few-shot speaker verification
 using the prototypical network,
 which is a metric learning method that is based on
 computing a distance against a mean embedding vector of a target class called prototype,
 and a threshold to verify the target speaker~\cite{wang2019centroid}.
Although these works achieve high performance of classification/verification from chunks of sounds as the input,
 to the best of our knowledge,
 any studies have not been done yet for few-shot detection from long continuous query sequence
 that contain not only the target events but also the other events and background noise.

\begin{figure}[t]
    \centering
    \centerline{\includegraphics[width=0.80\linewidth]{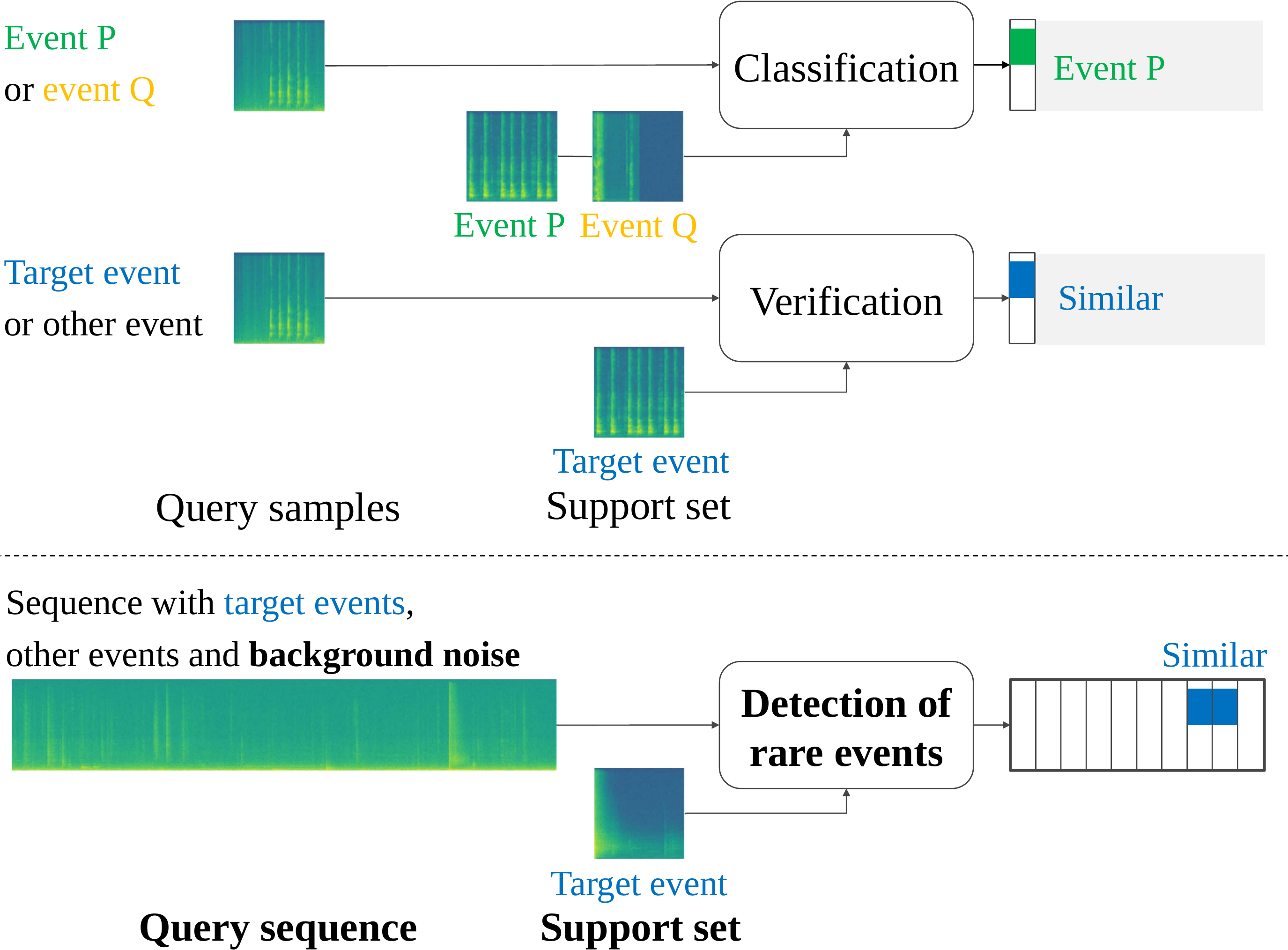}}
    \vspace{-3mm}
    \caption{Illustration of few-shot system overview.
             All few-shot systems take $k$-shot support sets,
             which means $k$ audio example(s) of target classes ($k=1$ in this figure).
             Unlike classification or verification systems,
             the detection system receives query sequence
             including target events, other events and background noise
             to estimate time-stamped activity.}
    \label{fig:overview}
    \vspace{-6mm}
\end{figure}

\begin{table*}[t]
    \centering
    \caption{Overview of deep metric learning tasks, loss functions, and sampling strategies.}
    \vspace{0mm}
        \begin{tabular}{l|lccrr} \toprule
        Paper                                           & Task                      & Fine-tuning   & Threshold & Loss or objective function& Sampling strategy         \\ \midrule
        Manocha~{\it et al.}~\cite{manocha2018content}  & retrieval                 & -             & -         & contrastive loss          & pair sampling             \\
        Pons~{\it et al.}~\cite{pons2019training}       & classification            & w/            & w/o       & cross-entropy loss        & episodic sampling         \\
        Koizumi~{\it et al.}~\cite{koizumi2019sniper}   & anomaly detection         & w/            & w/        & SNIPER objective          & anomalous sampling        \\
        Wang~{\it et al.}~\cite{wang2019centroid}       & few-shot verification     & w/ and w/o    & w/        & cross-entropy loss        & episodic sampling         \\
        \multirow{2}{*}{Chou~{\it et al.}~\cite{chou2019learning}} & \multirow{2}{*}{few-shot classification} & \multirow{2}{*}{w/o} & \multirow{2}{*}{w/o}
                                                                                                                & cross-entropy loss        & \multirow{2}{*}{episodic sampling} \\
                                                        &                           &               &           & w/ attentional similarity &                           \\ \midrule
        \multirow{2}{*}{Our work}                       & few-shot detection        & \multirow{2}{*}{w/o} & \multirow{2}{*}{w/}
                                                                                                                & contrastive loss          & balanced pair sampling    \\
                                                        & of rare events            &               &           & w/ weighted margin        & for background noise class\\ \bottomrule
        \end{tabular}
    \label{tb:overview_metric}
    \vspace{-6mm}
\end{table*}

Comparing to the few-shot classification/verification tasks given short query chunks as the input,
 which are shown in the upper half of Fig.~\ref{fig:overview},
 the few-shot detection tasks are given continuous samples including target events,
 other sound events and background noise,
 as in the lower half of Fig.~\ref{fig:overview}.
The detection tasks aim to estimate temporal activity of target events.
Therefore, it needs to be robust not to respond to the background noise.
However,
 the background noise class has not been considered explicitly in the most of few-shot learning works.
It is because only a few support examples of the background noise cannot cover all the variation of the noise.

In this paper,
 we propose metric learning with the explicit inclusion of background noise as an extra class
 for few-shot detection of rare sound events.
To prevent false positive reactions in the background noise sections,
 metric learning with the background noise class provides
 a feature space where the background noise class and event classes are sufficiently separated.
There are two contributions for learning such a feature space;
 a new sampling strategy and a loss function.
Various background noises are sampled during training with a controlled ratio,
 then, the loss function that weights the background noise class is designed
 to separate the target event classes and the noise class.
In the inference step,
 we calculate a temporal distance sequence
 between the learned embedding vectors of query sequence and target support examples,
 which is compared with a threshold.
Evaluations on few-shot detection tasks of rare sound events,
 using DCASE 2017 task2~\cite{mesaros2017dcase} and ESC-50~\cite{piczak2015esc},
 show that our proposed method outperforms metric learning without considering the background noise class.
The few-shot detection performance is also comparable to that of the DCASE 2017 task2 baseline system,
 which requires huge amount of annotated audio data for training.

\section{RELATED WORK}
\label{sec:related}

Deep metric learning methods
 aim to learn a feature space that retain the class neighborhood structure~\cite{wang2014learning,hoffer2015deep,koch2015siamese,vinyals2016matching,snell2017prototypical}.
The methods are used
 for retrieval tasks and few-shot classification tasks
 in various fields, such as computer vision and natural language processing.
The basic procedure of deep metric learning is as follows.
During training,
 sampled inputs are mapped to a feature space through an embedding network.
Then, the distances between the inputs in the feature space are calculated.
Lastly, the network parameters are updated based on a loss function using the distances.
In the inference step,
 by using the feature space, the input is classified to one of the target classes.

Regarding deep metric learning in acoustic signal processing~\cite{chou2019learning,wang2019centroid,manocha2018content,pons2019training,koizumi2019sniper,
                                                                   jansen2018unsupervised,sriskandaraja2018deep,thakur2019deep,turpault2019semi,lu2019class,monteiro2019combining,zhang2019few},
 we summarize an overview of tasks, loss functions, and sampling strategies in Table~\ref{tb:overview_metric}.
Manocha~{\it et al.} have worked on sound clip search task
 and used the contrastive loss~\cite{hadsell2006dimensionality},
 where a feature space is learned 
 based on a pair type that consists of the same class or different classes
 and a feature space distance~\cite{manocha2018content}.
They compare balanced and unbalanced sampling strategies
 for positive and negative label pairs.
Pons~{\it et al.} have tackled audio classification with few-data challenges~\cite{pons2019training}.
They follow the prototypical network~\cite{snell2017prototypical}
 and use the cross-entropy loss suitable for classification tasks.
The episodic sampling strategy~\cite{vinyals2016matching} is adopted during training,
 which samples only a few examples of each class as data points
 to simulate a few-shot classification scenario.
Koizumi~{\it et al.} have proposed
 that parameters of an anomaly detector are trained
 with an objective function called SNIPER,
 which suppresses the false negative rate of the overlooked anomaly,
 for a cascaded anomaly detection system~\cite{koizumi2019sniper}.
The system decided whether a machine sound clip contains anomaly or not.
Wang~{\it et al.} have tackled few-shot speaker verification
 using a threshold to verify the target speaker~\cite{wang2019centroid}.
They also focus on the prototypical network
 and compare the cross-entropy loss to the triplet loss.
Chou~{\it et al.} have assumed that an audio clip-level event classifier needs
 to be quickly adapted to recognize the new sound event without fine-tuning,
 i.e., support examples are not used to update network parameters~\cite{chou2019learning}.
They introduce an attentional similarity module,
 which guide the embedding network to pay attention to specific segments of
 an audio clip for classification of short or transient sound events.

This paper focuses on the few-shot detection task of rare sound events, which is also listed in Table~\ref{tb:overview_metric}.
We adopt the system without fine-tuning as in~\cite{chou2019learning},
 for the immediate response.
A threshold is used to detect only the target event class
 from a temporal sequence of distances computed by metric learning,
 which is commonly done in the verification tasks~\cite{wang2019centroid}.
To prevent false positive reactions in background noise sections,
 we propose a contrastive loss with weighted margin,
 and a balanced sampling strategy of background noise class,
 which provides a feature space where the background noise class and target event classes are separated.

\section{METHOD}
\label{sec:p_method}

\subsection{Basic processing flow of metric learning}
\label{ssec:basic}

\begin{figure}[t]
    \centering
    \centerline{\includegraphics[width=.80\linewidth]{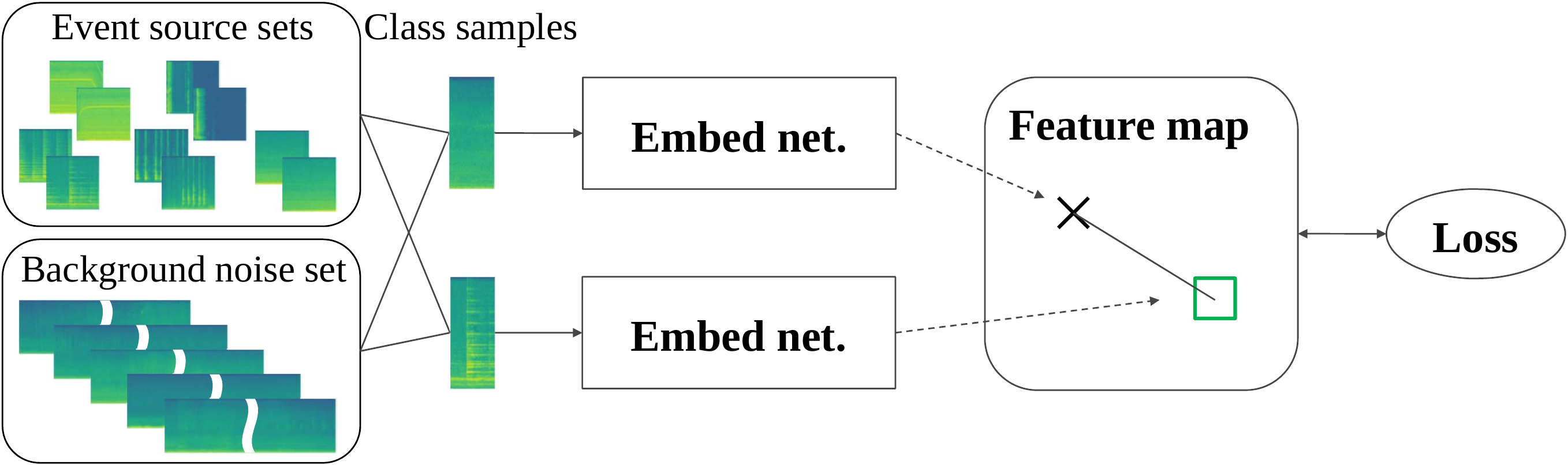}}
    \vspace{-3mm}
    \caption{In the training,
             pairs are mapped to a feature map via an embedding network
             whose parameters are updated with a loss function.}
    \label{fig:method_training}
    \vspace{-2mm}
\end{figure}

The goal is to train an embedding network
 that maps input samples to a feature space,
 where samples from the same class become closer
 while samples from different classes spread apart,
 as shown in Fig.~\ref{fig:method_training}.
Training consists of a pair sampling step and a training step for the embedding network.
In our work, we use convolutional neural network (CNN) architecture
 as the embedding network, $\mathcal{F}_{\mathrm{CNN},\theta}(\cdot)$,
 where $\theta$ is the parameters of the network.

In the sampling step,
 let $\Vec{X}_{1}, \Vec{X}_{2}\in{\mathbb{R}}^{F \times T}$ be training samples
 consisting of $F$-dimensional $T$ frame acoustic features
 and let $y_{1}, y_{2}\in\{ 1, \dots, {C}_{\mathrm{train}} + 1 \}$
 be class labels assigned to the samples,
 where ${C}_{\mathrm{train}}$ is the total number of classes in event source sets.
The other one is the background noise class.
The training sample set consists of
 event class samples and background noise class samples.
Event class samples are generated
 by mixing samples from the event source sets and samples from the background noise set
 with random event-to-background ratios (EBR).
Background noise class samples are randomly selected only from the background noise set.

The parameters of the network are updated
 by minimizing the following objective function
 in a pair consisting of $\Vec{X}_{1}$ and $\Vec{X}_{2}$:
\begin{align}
    \min_{\theta} \sum_{\Vec{X}, y}
     \mathcal{L}(\mathcal{D}_{\theta}(\Vec{X}_{1}, \Vec{X}_{2}), y_{1}, y_{2}),
    \label{eq:objective_function}
\end{align}
where $\mathcal{L}$ is a loss function,
 and $\mathcal{D}_{\theta}$ is the Euclidean distance in the feature space:
\begin{align}
    \mathcal{D}_{\theta}(\Vec{X}_{1}, \Vec{X}_{2})
     = || \mathcal{F}_{\mathrm{CNN},\theta}(\Vec{X}_{1}) - \mathcal{F}_{\mathrm{CNN},\theta}(\Vec{X}_{2}) ||.
    \label{eq:distance_function}
\end{align}
Since we focus on the pair that consists of the same class or different classes,
 we adopt the contrastive loss function defined in~\cite{hadsell2006dimensionality} as base.
Let $l$ be the label assigned to the pair.
$l = 1$ if the inputs belong to the same class ${y}_{1} = {y}_{2}$;
 otherwise $l = 0$.
Then, the loss function $\mathcal{L}$ is defined as
\begin{align}
       &\mathcal{L}(\mathcal{D}_{\theta}(\Vec{X}_{1}, \Vec{X}_{2}), y_{1}, y_{2}) \nonumber \\
     =& {l} \mathcal{D}_{\theta}(\Vec{X}_{1}, \Vec{X}_{2}) ^ 2 
     + (1 - {l}) \max ({m} - \mathcal{D}_{\theta}(\Vec{X}_{1}, \Vec{X}_{2}), 0) ^ 2,
    \label{eq:contrastive_loss}
\end{align}
in which $m > 0$ is a margin.
The network parameters $\theta$ are trained
 so that the pairs are closer if the class label $l = 1$
 and are farther apart if $l = 0$.

\begin{figure}[t]
    \centering
    \centerline{\includegraphics[width=.80\linewidth]{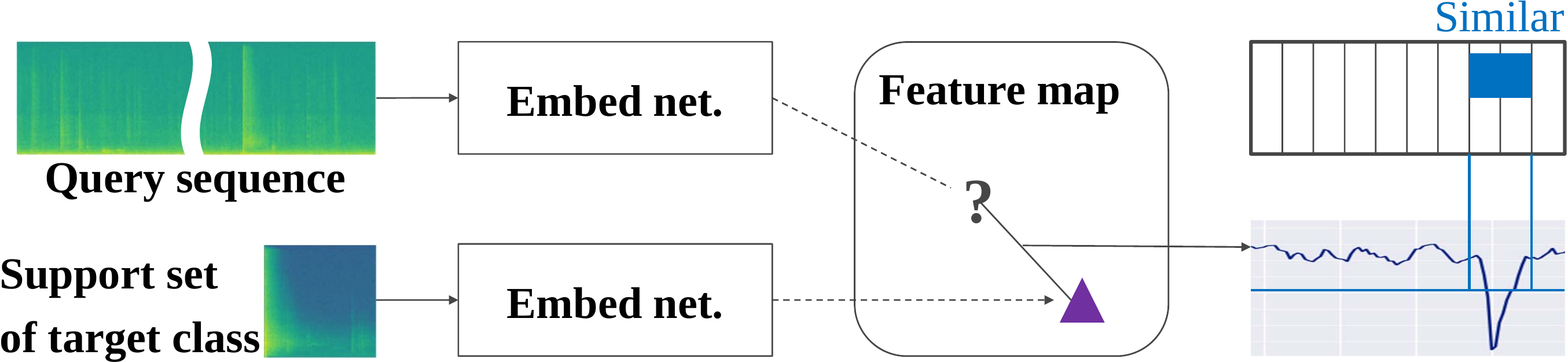}}
    \vspace{-3mm}
    \caption{In the inference,
             the distance between the support set and the query sample in the feature map
             is calculated and compared with a threshold to decide
             whether the query sample is in the target event class or not.}
    \label{fig:method_inference}
    \vspace{-5mm}
\end{figure}

The inference step is summarized in Fig.~\ref{fig:method_inference}.
Given a $k$-shot support set
 $\mathcal{S} = \{ \Vec{S}_{i}\in{\mathbb{R}}^{F \times T} ({i} = 1, \dots, k) \}$,
 which consists of $k$ examples from the target event class not seen during training.
$k$ is typically a small number from 1 to 10.
Let $\Vec{Q}_{n}\in{\mathbb{R}}^{F \times T}$ be a query sample in time frame $n$,
 which is a part of long continuous audio sequence
 including target events, other events, and background noise.

A distance space between the support set and the query sample,
 $\mathcal{D}_{\theta}(\mathcal{S}, \Vec{Q}_{n})$,
 is calculated as follows:
\begin{align}
    \mathcal{D}_{\theta}(\mathcal{S}, \Vec{Q}_{n})
    = || \Vec{\mu}_{k} - \mathcal{F}_{\mathrm{CNN},\theta}(\Vec{Q}_{n}) ||,
    \label{eq:distance_function_prototype}
\end{align}
where $\Vec{\mu}_{k}$ is a mean vector of the embedded support examples called prototype~\cite{snell2017prototypical},
 $\Vec{\mu}_{k} = \frac{1}{k} \Smallsum_{\Vec{S}_{i}\in\mathcal{S}} \mathcal{F}_{\mathrm{CNN},\theta} (\Vec{S}_{i})$.
The distance $\mathcal{D}_{\theta}(\mathcal{S}, \Vec{Q}_{n})$ is compared with a threshold $\sigma$
 to decide whether the query sample is the target event class or not.
Finally, the classification result is output
 alongside with the time-stamp of the partial input of the audio sequence.

\subsection{Balanced sampling strategy for background noise class}
\label{ssec:sampling}

\begin{figure}[t]
    \centering
    \centerline{\includegraphics[width=.60\linewidth]{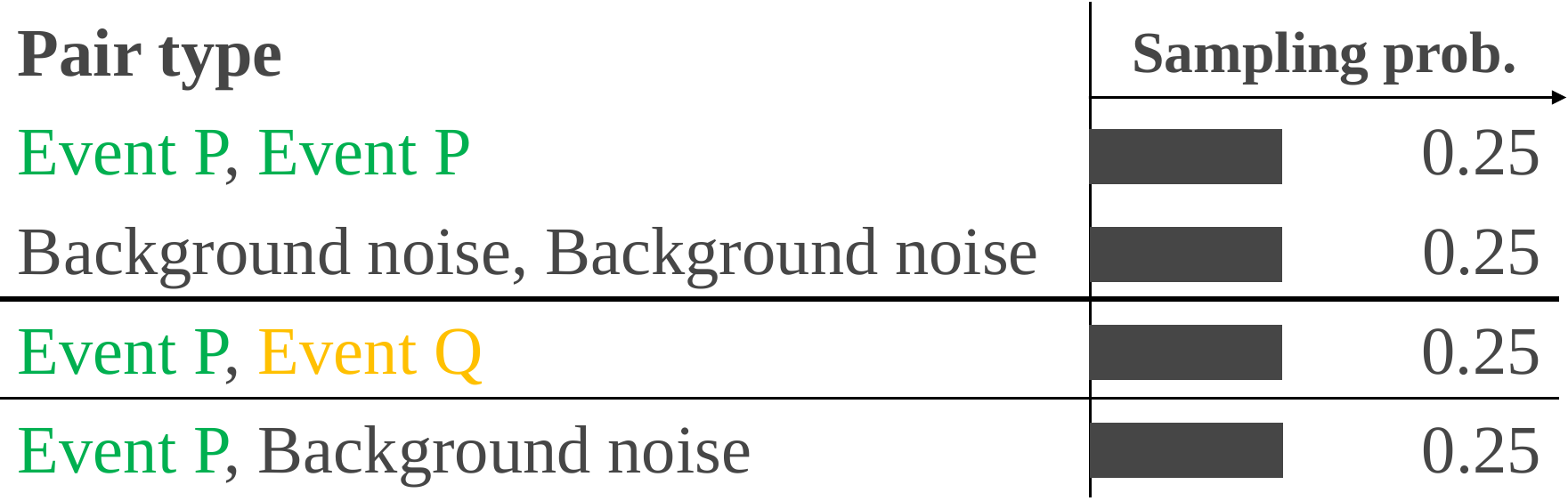}}
    \vspace{-4mm}
    \caption{In the proposed sampling strategy,
             half of the pairs contain background noise class.}
    \label{fig:method_sampling}
    \vspace{-3mm}
\end{figure}

We aim to select samples from sound event classes and background noise class
 with a suitable ratio for the detection task.
To separate the background noise class and the sound event classes in the feature space,
 the background noise class is required to be sampled
 at the same ratio as that of the other event classes all together in Fig.~\ref{fig:method_sampling}.
In our preliminary experiment,
 when the background noise class is sampled at the same ratio as each other event class,
 the background noise class is almost missed in the training step
 and is not sufficiently separated from the sound event classes.

\subsection{Contrastive loss with weighted margin}
\label{ssec:loss}

\begin{figure}[t]
    \centering
    \centerline{\includegraphics[width=.95\linewidth]{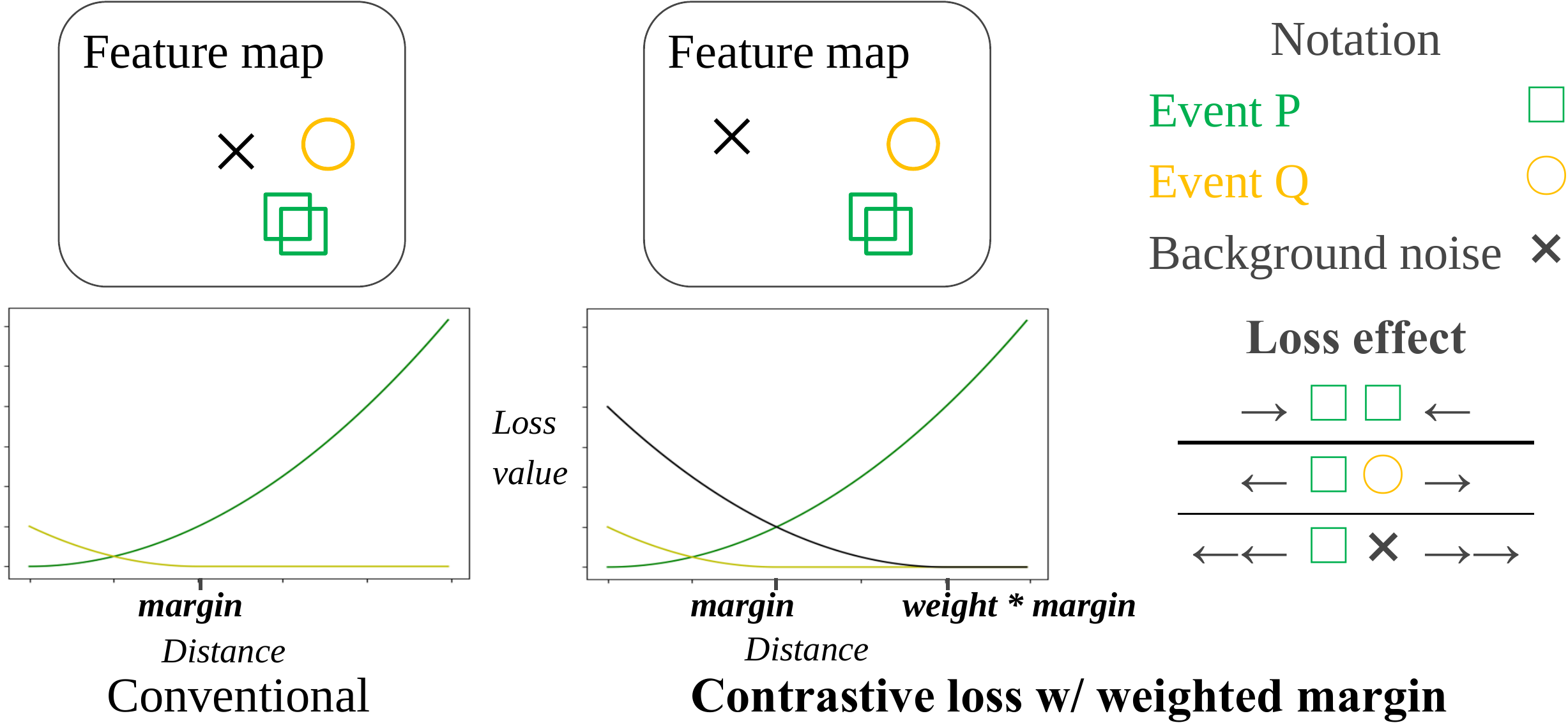}}
    \vspace{-3mm}
    \caption{A weight for the margin is introduced into the contrastive loss
             to separate the background noise class far away from event classes.}
    \label{fig:method_loss}
    \vspace{-5mm}
\end{figure}

We introduce a new weighting algorithm into the conventional contrastive loss~\cite{hadsell2006dimensionality},
 by applying a large weight to the background noise class to be far from the sound event classes in the feature space.
We define $w$ as a weight for the margin
 into the contrastive loss function.
$w > 1$ if the inputs have background noise class; otherwise $w = 1$
 to separate the background noise class far away from event classes.
\begin{align}
       &\mathcal{L}(\mathcal{D}_{\theta}(\Vec{X}_{1}, \Vec{X}_{2}), y_{1}, y_{2}) \nonumber \\
     = &{l} \mathcal{D}_{\theta}(\Vec{X}_{1}, \Vec{X}_{2}) ^ 2 
     + (1 - {l}) \max ({w}{m} - \mathcal{D}_{\theta}(\Vec{X}_{1}, \Vec{X}_{2}), 0) ^ 2,
    \label{eq:contrastive_loss4noise}
\end{align}
where the weighted margin ${w}{m}$ in the second term
 increases when the input pair consists of a background noise class sample and an event class sample
 to separate them from each other in the feature space, as in Fig.~\ref{fig:method_loss}.
The weighting makes
 the background noise class farther away than other event classes, from each event class.

Looking only at pairs of event classes,
 they follow the usual contrastive loss~\cite{hadsell2006dimensionality}.
An effect of pairs of a background noise class sample and an event class sample is
 that each event class is expected to be separated almost equally far from the background noise class
 when training has progressed sufficiently.
If the dimensionality of the feature space is large enough,
 mapping of the sound event classes is expected to be learned as usual,
 and the background noise class is expected to be mapped away from the event class regions.

\begin{table*}[t]
    \centering
    \caption{Few-shot detection performances of rare sound events (event-based F1 score)
             on evaluation tasks.
             The query sequence sets consist of 30-second audio clips
             from DCASE 2017 task2 development and evaluation sets.}
    \vspace{0mm}
        \begin{tabular}{lr|cccc|cccc} \toprule
                                                    &           & \multicolumn{4}{c|}{Development set}          & \multicolumn{4}{c}{Evaluation set}            \\
        Method                                      & Support   & Baby.     & Glass.    & Gun.      & Avg.      & Baby.     & Glass.    & Guns.     & Avg.      \\ \midrule
        Metric learning w/o background noise class  & 10-shot   & 0.23      & 0.19      & 0.08      & 0.17      & 0.21      & 0.14      & 0.07      & 0.14      \\
        + background noise detector                 & 10-shot   & {\bf 0.53}& 0.41      & 0.47      & 0.47      & {\bf 0.59}& 0.50      & 0.25      & 0.45      \\ \midrule
        Metric learning w/ background noise class   & 1-shot    & 0.33      & 0.35      & 0.25      & 0.31      & 0.35      & 0.28      & 0.27      & 0.30      \\
        (proposed)                                  & 5-shot    & 0.42      & 0.61      & 0.47      & 0.50      & 0.45      & 0.55      & 0.44      & 0.48      \\
                                                    & 10-shot   & 0.45      & {\bf 0.66}& {\bf 0.56}& {\bf 0.56}& 0.49      & {\bf 0.58}& {\bf 0.47}& {\bf 0.51}\\ \midrule
        \multicolumn{2}{c|}{DCASE 2017 task2 baseline} 
        & \multirow{2}{*}{(0.72)} & \multirow{2}{*}{(0.89)} & \multirow{2}{*}{(0.57)} & \multirow{2}{*}{(0.73)} & \multirow{2}{*}{(0.67)} & \multirow{2}{*}{(0.79)} & \multirow{2}{*}{(0.47)} & \multirow{2}{*}{(0.64)} \\
        \multicolumn{2}{c|}{trained with 500 mixture audio examples~\cite{mesaros2017dcase}} & & & & & & & & \\ \bottomrule
        \end{tabular}
    \label{tb:results_-6to6}
    \vspace{-6mm}
\end{table*}

\section{EXPERIMENTAL EVALUATION}
\label{sec:exp}

\subsection{Experimental settings}
\label{ssec:tasks}

We evaluated the proposed method using few-shot detection experiments of rare sound events.
DCASE 2017 task2~\cite{mesaros2017dcase} and ESC-50~\cite{piczak2015esc}
 were used for the detection tasks.
DCASE 2017 task2 is a detection task of rare sound events
 using target sound events (babycry, glassbreak, and gunshot)
 and 30-second background noise clips from TUT Acoustic Scenes 2016 dataset~\cite{mesaros2016tut}.
A synthesizer provided in the challenge
 was used to generate mixtures of target events and background noise clips.
The ESC-50 contains 2,000 5-seconds audio clips labeled with 50 classes,
 each having 40 examples.
The sound categories cover sounds of animals, natural sounds, human sounds,
 and other interior/exterior characteristic sounds.
The background noise set for training was from DCASE 2017 task2 devtrain background audio.
The event source sets were 43 classes,
 with labels from 0 to 44 of ESC-50 excluding babycry and glassbreak classes,
 since they were target event classes in evaluation step.
The rest 5 classes were used to verify metric learning training through their loss value.
In the evaluation step,
 query sequence sets were from DCASE 2017 task2 development and evaluation sets,
 following its setup.
Each set consisted of 30-second generated audio clips.
There were 500 clips per each target sound event.
The event presence rate was 0.5
 (250 clips with target event present and 250 clips of only background).
The EBR for each mixed clip was randomly selected from -6, 0, or 6 dB.
Few-shot support set for each target class was from DCASE 2017 task2 devtrain target sound events.
There were about 100 target sound events for each class.
We randomly selected $k$ target events as $k$-shot support examples: $k \in {1, 5, 10}$.
Since results of the experiment might vary depending on which support examples were selected~\cite{pons2019training},
 we carried out each experiment 20 times per fold of data, and reported averaged scores across the folds.
We used an event-based F1 score metric~\cite{mesaros2016metrics},
 and the metric was calculated using onset-only condition with a collar of 500 ms.

Sampling frequency was standardized at 16 kHz in our experiments.
Acoustic feature was $F = 40$-channel log Mel-scale filter bank per each frame
 where short-time Fourier transform was applied with configurations of 30 ms frame length and 10 ms frame overlap.
In the training step,
 each sample was generated on-the-fly~\cite{erdogan2018investigations}.
Each sample consisted of $T = 100$ frame acoustic features.
The class $y$ of each sample is a one-hot vector focused on central 20 frames.
When the pair for the contrastive loss was created,
 first, the label $l$ in~(\ref{eq:contrastive_loss4noise}) is selected from 0 or 1,
 then the class $y$ is selected randomly for each of the pair.
The EBR for each mixed sample of pairs was randomly selected from 0 to 18 dB.
According to the sampling strategy in section~\ref{ssec:sampling},
 half of the pairs contained background noise class.
The ratio of the cases of $l=0$ and $l=1$ was evenly set as 0.5.
Further we divided the ratio in each case evenly.
Specifically, in the case of $l=1$,
 the occurrence probability of pairs with only background noise class was set to 0.25,
 and that of pairs with only the same sound event class was set to 0.25.
Similarly, in the case of $l=0$, 
 the probability of pairs with the background noise class and a sound event class was set to 0.25,
 and that of pairs with different sound event classes was set to 0.25.
Each sound event class was evenly assigned.
The embedding network was a lightweight version of ResNet architecture~\cite{he2016deep},
 which has approximately 70 k parameters,
 and mapped each sample to a sufficiently large 128-dimension feature space.
The margin of the loss function $m$ was set to 1,
 and the weight coefficient for the margin was set as $w = 2$ if the inputs have background noise class;
 otherwise $w = 1$.
The Adam optimizer is used with $0.001$ learning rate~\cite{kingma2014adam}.
We use a batch size of 64 and set the maximal number of epochs to 100 based on the event source sets.

As a support example of each target sound event in the inference step,
 we used features for 100 frames centered on each event onset time.
The distance $\mathcal{D}_{\theta}$ was calculated in every 20 frames.
The threshold $\sigma$ was optimized using the development set
 for each method and each target sound event class.
We used metric learning without background noise class
 as a baseline for comparison.
Since the baseline method alone was not robust
 enough to prevent false positive errors in background noise sections,
 the method was combined with a detector
 designed to identify background noise section and eliminate them in the inference step.
Oracle detector was not used
 because it was difficult to define the other event sections in this case.
The detector was based on a supervised binary classifier
 with the same network configuration and training data as the proposed method.
In our preliminary experiments,
 the detector performed at a recall of about 0.6 in the same development set.

\subsection{Experimental results}
\label{ssec:res}

The event-based F1 scores are listed in Table~\ref{tb:results_-6to6},
 where performance of few-shot detection of rare sound events were evaluated,
 using query sequence sets of 30-second generated audio clips
 from DCASE 2017 task2 development and evaluation sets.
Compared with the baseline method combined with the background noise detector,
 the proposed method showed 0.06 pt higher average F1 score in evaluation set.
In addition,
 the proposed method was also simpler system since it is based on only metric learning.
The performance of proposed method improved as the number of shots $k$ increased.
As the long continuous query sequence
 included target events, other events, and various background noise,
 metric learning without background noise class
 was vulnerable to various background noise.
Overall,
 the low F1 scores indicated the difficulty of the few-shot detection task of rare sound events.
The best event-based F1 score for the gunshot class, 0.47,
 was comparable to the 0.47 score of the DCASE 2017 task2 baseline system,
 which was trained with sizable amounts of annotated sound data~\cite{mesaros2017dcase}.

\section{CONCLUSION}
\label{sec:conclusion}

We proposed metric learning with background noise class
 for few-shot detection of rare sound events.
To prevent false positive reactions to the input sequence,
 the metric learning provided a feature space
 where the background noise class and event classes are sufficiently separated.
We designed a new sampling strategy,
 where various background noises were sampled during training with a controlled ratio.
Then,
 the loss function that weights the background noise class was designed
 to spread between the event classes and the noise class.
In the inference step,
 we calculated a distance sequence between the learned embedding vectors of query sequence and target support examples,
 which is compared with a threshold.
Evaluations on few-shot detection task of rare sound events,
 using DCASE 2017 task2~\cite{mesaros2017dcase} and ESC-50~\cite{piczak2015esc},
 showed that our proposed method outperformed metric learning without considering the background noise class.
The few-shot detection performance was also comparable to
 that of the DCASE 2017 task2 baseline system,
 which requires huge amount of annotated audio data for training.

\bibliographystyle{IEEEtran}
\bibliography{refs}

\end{document}